# Nonreciprocity in Magnon Mediated Charge-Spin-Orbital Current Interconversion

José Omar Ledesma-Martin, Edgar Galindez-Ruales, Sachin Krishnia,* Felix Fuhrmann, Minh Duc Tran, Rahul Gupta, Marcel Gasser, Dongwook Go, Akashdeep Kamra, Gerhard Jakob, Yuriy Mokrousov, and Mathias Kläui*

**ABSTRACT:** In magnetic systems, angular momentum is carried by spin and orbital degrees of freedom. Nonlocal devices, comprising heavy-metal nanowires on magnetic insulators like yttrium iron garnet (YIG), enable angular momentum transport via magnons. These magnons are polarized by spin accumulation at the interface through the spin Hall effect (SHE) and detected via the inverse SHE (iSHE). The processes are generally reciprocal, as demonstrated by comparable efficiencies when reversing injector and detector roles. However, introducing Ru, which enables the orbital Hall effect (OHE), disrupts this reciprocity. In our system, magnons polarized through combined SHE and OHE and detected via iSHE are 35% more efficient than the reverse process. We attribute this nonreciprocity to nonzero spin vorticity, resulting from varying electron drift velocities across the Pt/Ru interface. This study highlights the potential of orbital transport mechanisms in influencing angular momentum transport and efficiency in nonlocal spintronic devices.

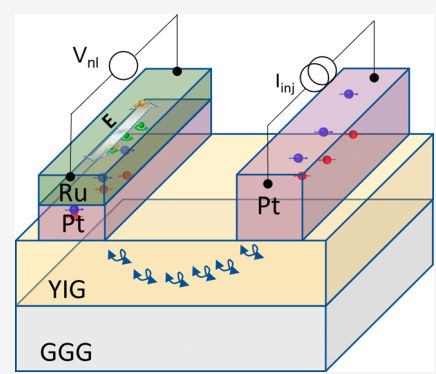

**KEYWORDS:** Orbital torques, Spin-orbitronics, Nonlocal magnon detection, Orbital Hall effect

Spin current generation/detection mechanisms and spin transport in magnetic systems have been a key research area over the past decade due to their possible applications and the fundamental understanding of these mechanisms.[1,2] The spin Hall effect (SHE) in the bulk of the heavy metal[3−6] and the Rashba-Edelstein effect (REE)[7−9] at inversion asymmetric interfaces involving high spin–orbit interactions, have been intensively investigated to generate spin currents and a nonequilibrium spin density from a charge current. The spin current and spin accumulation may diffuse and exert a spin–orbit torque on an adjacent magnetic layer.[2,4,6,10,11] The reciprocal process involves spin currents generated by magnetization precession, diffusing into a heavy-metal layer and detected as a voltage via the inverse SHE (iSHE) and/or inverse REE (iREE) effects.[12−15] Notably, the direct (SHE, REE) and inverse (iSHE, iREE) spin-current conversion efficiencies have been found to be similar to each other. Devices studied include nonlocal devices, where spin angular momentum transfer differs based on whether magnetic layers are conducting or insulating.[15−17] In metallic ferromagnets, spin transport involves conduction electrons,[6] while insulating magnets (e.g., YIG/Pt structures) require spin transport through magnon creation and annihilation.[17−20]

Spin current-induced magnon dynamics in insulating magnets interfaced with Pt have been widely studied over the past decade.[18,20,21] In particular, YIG, a magnetic insulator, known for its low magnetic damping[22] and long-distance magnon propagation,[17] is ideal for magnonic studies. The interaction of YIG with spin currents injected from or into thin Pt nanowires generates and detects magnons electrically via nonlocal measurements, bypassing the need for charge flow within the insulator. This spin current creates spin accumulation at the YIG interface, manipulating its magnetic states through spin angular momentum transfer to control magnon generation and propagation. The inverse process, where magnons from YIG are converted back into electrical signals in Pt via the iSHE, exhibits crucial bidirectional reciprocity of the conversion.[17]

Recent experimental studies and theoretical models have revealed that charge currents can induce orbital accumulation, generating transverse orbital currents via the orbital Hall and orbital Rashba-Edelstein effects (OHE and OREE),[23−27] analogous to spin counterparts. These effects stem from orbital textures in momentum space and can occur without spin–orbit coupling (SOC).[28] Such phenomena have been predicted in materials like transition metals,[29−33] metal-oxide

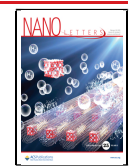

3247

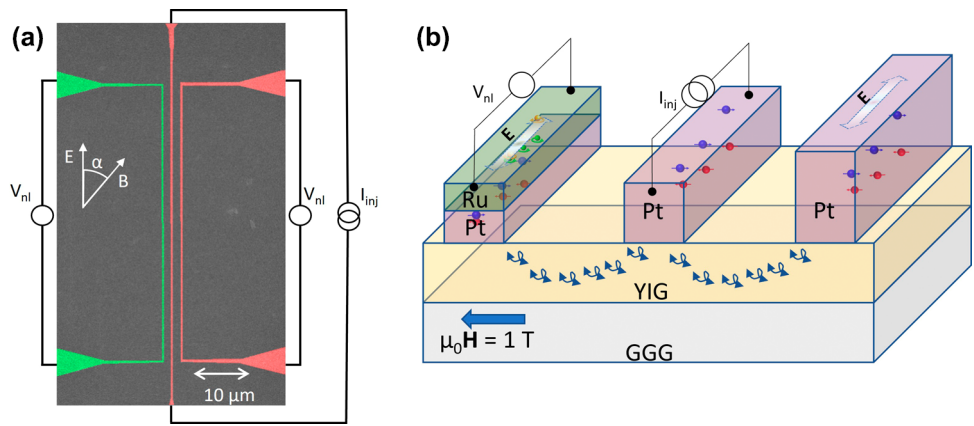

**Figure 1.** (a) Colored scanning electron microscope (SEM) micrograph of a nonlocal magnon transport device with the measurement scheme. The device consists of three parallel wires. The two Pt wires having thicknesses of 7 nm and 13 nm (center and right, in red) serve as a source and/or detector of spin current, and the third Pt/Ru wire (left, green) can generate and detect spin and orbital currents. The spacing between the two wires varies from 500 nm to 2 μm in different devices. The angle between the direction of the charge current (parallel to the wire) and the external magnetic field is represented by $\alpha$. (b) Schematic representation of the cross-section of a device with the measurement scheme. A spin accumulation is generated by a charge current injection in the middle Pt wire due to the SHE, as shown by red and blue electrons. This spin accumulation generates magnons via spin angular momentum transfer to YIG, which are then converted into spin current and a resultant voltage at the detector Pt (or Ru) wire due to the iSHE (or iSHE + iOHE).

interfaces,[34,35] and two-dimensional systems.[36−38] The emerging field of orbitronics[23] explores orbital angular momentum to harness orbital currents independently or with the spin degree of freedom. This interest has spurred efforts to observe orbital transport phenomena. Additionally, orbital currents convert efficiently into spin currents in high-spin−orbit coupling materials, enabling ferromagnet manipulation. This phenomenon is termed orbital torques. Recent findings of the inverse OREE (iOREE) at interfaces like LaAlO$_3$/SrTiO$_3$[39] and Pt/CuO$_x$[40] show orbital currents convert into charge currents, boosting the potential of nonlocal magnon transport devices and broadening their applications. These developments allow for comparisons between OHE, OREE, iOHE, and iOREE efficiencies in magnon transport devices on insulating magnets.

While approximate reciprocity in spin-charge interconversion has been reported,[15] some studies claim large orbital torques can arise in systems with weak orbital pumping. Previous studies have failed to compare orbital-to-charge and charge-to-orbital conversion in identical samples, where it is unclear if the differences come from sample variations or intrinsic nonreciprocity. To resolve this, it is essential to study these effects in a single device.

This work investigates the role of a strong OHE material (Ru) in the reciprocity of magnon generation and detection in a Ru/Pt-based nonlocal device. Specifically, we analyze magnon generation via OHE and SHE in a Ru/Pt wire and detection via the iSHE in Pt, as well as the reciprocal process: magnon generation via SHE in Pt and detection via iOHE and iSHE in the Ru/Pt wire. We identify the differences and attribute them to charge-to-orbital interconversion in the Ru layer and its interface with Pt.

To examine the reciprocity between magnon generation and detection, we study magnon transport in a nonlocal geometry, as shown in Figure 1(a). Our device comprises two Pt wires and one Pt/Ru wire, all deposited on a 1.5 μm thick YIG film. YIG's low damping enables long magnon propagation lengths,[17] Additionally, its electrically insulating nature mitigates electrical shunting effects and simplifies the magnon generation complexity associated with spin and/or orbital currents originating from the self-torque mechanism.

The YIG thin films were grown by liquid phase epitaxy on GGG substrates, provided by Matesy GmbH. Electron beam lithography was used to define the device pattern. After fabricating alignment markers, two parallel Pt wires (250 nm wide, 7 nm, and 13 nm thick, respectively) are fabricated. We use a Singulus Rotaris DC magnetron sputtering tool for the metal deposition. We then fabricate a third wire, also 250 nm wide and parallel to the Pt wires with equal spacing. A Pt(1.5 nm)/Ru(4 nm)/MgO(2 nm)/Ta(2 nm) multilayer stack was deposited for orbital current generation and detection. Ru generates and detects orbital currents, while Pt converts these into spin currents via spin−orbit interaction. The MgO/Ta layer protects Ru from oxidation. The device comprises three parallel wires (250 nm wide, 50 μm long) separated by 500 nm to 2 μm. A typical scanning electron microscopy image of the device is shown in Figure 1(a).

To characterize magnon generation and detection, we conduct nonlocal electrical measurements. We inject current into one Pt wire (the injector wire), which generates a transversally polarized spin current and a spin accumulation at the YIG/Pt interface, as illustrated in Figure 1(b). This spin accumulation can couple to the YIG magnetization via the exchange interaction, thereby polarizing magnons within the YIG; this mechanism allows the spin currents to polarize magnons, but not the orbital currents, which have to go through an orbital-spin interconversion via SOC before interacting with the magnetization. These magnons then propagate toward the other Pt electrode (the detector electrode). At the detector electrode, a reciprocal process occurs: the magnons interact with the Pt and are converted into a spin current at the interface, which in turn induces a voltage signal due to the iSHE in Pt. For each measurement, the device is placed in a cryostat with an in-plane 0.1 T rotating magnetic field. We used a Keithley 6220 as a current source in conjunction with a Keithley 2182A nanovoltmeter in Delta mode, and a constant pseudo-DC current was applied to the injector wire, and the corresponding voltage at the detector wire was measured using a Keithley 2182A nanovoltmeter. The Delta mode measures the voltage for pulses of the same amplitude but opposing polarity, eliminating any thermal offset signals and being equivalent to first harmonic lock-in measurements. We define a nonlocal resistance ($R_{nl}$) of the detector wire as

$$R_{nl} = \frac{V_{det}}{I_{inj}} \quad (1)$$

where $V_{det}$ is the voltage measured at the detector wire, and $I_{inj}$ is the current applied to the injector wire. Given the symmetry constraints of the SHE and iSHE mechanisms, the magnon injection/detection process only occurs when there is a component of the spin accumulation that is collinear with the magnetization.[17] Therefore, we measure the dependence of $R_{nl}$ on the angle between the magnetization and current direction ($\alpha$) by placing the sample in an 0.1 T in-plane rotating field (greater than the anisotropy field of YIG).

In Figure 2(a), we present the dependence of $R_{nl}$ as a function of $\alpha$ for a fixed $I_{inj}$ = 0.3 mA, in a device with a gap of

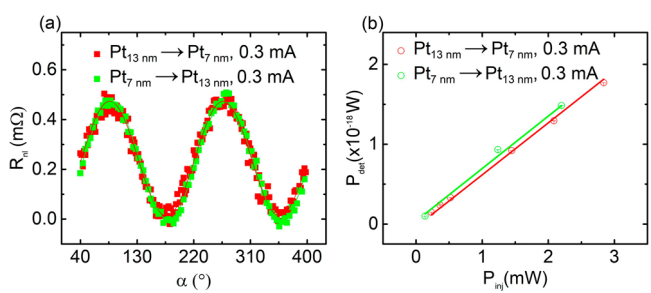

**Figure 2.** (a) Nonlocal resistance ($R_{nl}$) as a function of the angle ($\alpha$) between the charge current and the magnetization direction. The red data points correspond to the current injected in the 13 nm thick Pt wire and nonlocal voltage measurement in the 7 nm thick Pt wire ($Pt_{13\ nm} \rightarrow Pt_{7\ nm}$). Whereas the green points correspond to the current injected in the 7 nm thick Pt wire and nonlocal voltage measurement in the 13 nm thick Pt wire ($Pt_{7\ nm} \rightarrow Pt_{13\ nm}$). Solid lines represent the sinusoidal fit. An offset baseline has been removed so the nonlocal resistance values are zero at $\alpha$ = 0 degrees. (b) The detected power $P_{det} = V_{det}^2/R_{det}$ (where $R_{det}$ is the resistance of the detector wire) as a function of injected power $P_{inj} = I_{inj}^2 R_{inj}$ (where $R_{inj}$ is the resistance of the injector wire) for a device with 1 $\mu$m spacing between the two wires.

1 $\mu$m between 13 nm and 7 nm thick Pt wires. Note that the current also induces Joule heating, which can generate magnons; however, in this work, we focus only on magnons generated by spin currents by measuring the odd components of voltages for two different current polarities in delta mode. As expected, the $R_{nl}$ shows a typical $\sin(2\alpha)$ dependence (up to an offset and a scaling constant) in contrast to thermal magnons, which show a $\sin(\alpha)$ dependence (up to an offset and a scaling constant).[41] It shows that the maximum amplitude corresponds to the generation/detection of magnons when the spin polarization and magnetization are collinear and the minimum when they are perpendicular. Next, we verify the reciprocity of magnon transport by interchanging the current and voltage contacts of the injector and detector wires. For identical-thickness Pt wires, we find the expected reciprocity, as shown in the Supporting Information (S1). To make sure that the reciprocity is not only found for nominally identical wires, we use Pt wires with different thicknesses and resistances: 13 and 7 nm, with resistances of 5 k$\Omega$ and 13 k$\Omega$, respectively, as injector and detector. This approach also allows us to rule out the possibility that the effect is only observed for certain resistance combinations.

The measured $R_{nl}$ for the two directions, $Pt_{7\ nm} \rightarrow Pt_{13\ nm}$ (in green) and $Pt_{13\ nm} \rightarrow Pt_{7\ nm}$ (in red), as well as sinusoidal fits for $R_{nl}$, is shown in Figure 2(a). Here, the subscripts 7 nm and 13 nm stand for the thickness of the Pt wires to distinguish them from each other. From the measurement and corresponding fits, we find $R_{nl}$ ($Pt_{7\ nm} \rightarrow Pt_{13\ nm}$) $\approx R_{nl}$ ($Pt_{13\ nm} \rightarrow Pt_{7\ nm}$) = 451 $\mu\Omega \pm$ 5.4 $\mu\Omega$. As $R_{nl}$ exhibits similar amplitudes within the error bars in both measurements, we thus conclude that reciprocity is maintained between magnon generation and detection via the SHE and iSHE in our Pt $\rightarrow$ Pt device.

Additionally, we perform these measurements for several injected currents and compare the injected power with the detected power between the two Pt wires of different resistances. As shown in Figure 2(b), the detected power increases linearly with the injected power and it exhibits a similar slope (6.51 × 10$^{-16}$ ± 2.6 × 10$^{-18}$ for $Pt_{13\ nm} \rightarrow Pt_{7\ nm}$ and 6.57 × 10$^{-16}$ ± 4.47 × 10$^{-17}$ for $Pt_{7\ nm} \rightarrow Pt_{13\ nm}$) when the injector and detector are interchanged. Based on this observation, we adopt a power-to-power comparison approach hereafter.

Furthermore, the measurements on identical Pt wires (in terms of thickness and resistance) further confirm that magnon generation and detection are reciprocal processes. The details are provided in the Supporting Information (S1).

Next, we analyze whether the reciprocity holds for magnon generation/detection when the magnon current is generated via the recently discovered OHE mechanisms. To this end, we replaced one of the 13 nm thick Pt wires with an orbital current generation electrode, i.e., the Ru wire. Note that the spin-charge interconversion processes in the Ru wire exhibits a dominating OHE as demonstrated in our previous work.[30,42] We expect the generation of orbital currents in Ru and its efficient conversion to a spin current in the adjacent 1.5 nm Pt layer. The resulting spin accumulation at the interface with YIG polarizes the magnons, which are detected at the 7 nm Pt detector wire as a voltage generated via the iSHE mechanism, as explained previously. In Figure 3(a), we show the $R_{nl}$ dependence as a function of $\alpha$ for a current $I_{inj}$ = 0.2 mA in injector Ru wire (blue circle). The $R_{nl}$ shows the expected angular dependence with an amplitude $R_{nl}$ = 2.7 m$\Omega$ ± 17 $\mu\Omega$. Next, we reverse the position of the injector and detector, i.e., we inject the current into the Pt wire, detect the voltage in the Ru wire, and repeat the same measurement, as shown in Figure 3(a) (red). The magnons are polarized by the spin accumulations at the YIG/Pt interface and propagate toward the Ru electrode. These magnons accumulate underneath the Ru wire and create a spin current that flows into the 1.5 nm Pt (below the Ru layer) and is converted into an orbital current. As the thickness of Pt is smaller than the orbital diffusion length, the converted orbital current further diffuses into the Ru layer, where it is converted into a voltage via the inverse OHE; at the same time, the Pt layers are in the range of the spin diffusion length (1.8−2.5 nm in similar systems)[27,30] ensuring high efficiency of spin to orbital conversion and avoiding the injection of spin current in the Ru layer. The amplitude of $R_{nl}$ for magnon injection via pure SHE and detection via iSHE and iOHE is found to be $R_{nl}$ = 2.1 m$\Omega$ ± 28 $\mu\Omega$ for an injected current $I_{inj}$= 0.3 mA in the Pt electrode.

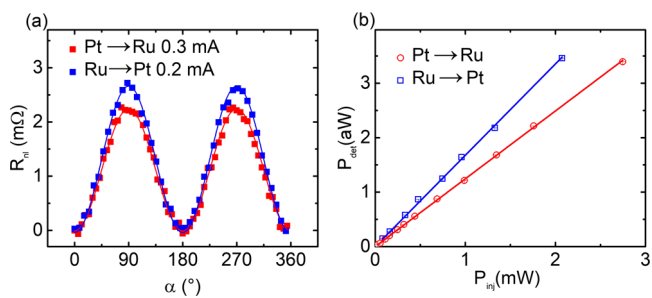

**Figure 3.** (a) Comparison of the nonlocal resistance as a function of the angle between the injected current and magnetization direction, when current is injected into Pt (Ru) wire and voltage is detected in Ru (Pt) wire, depicted by red circles (blue squares). The injected current is adjusted to account for differences in the resistance of the two electrodes by maintaining the same injected power at approximately 1.1 mW. (b) Linear fits for power injected ($P_{inj}$) to power detected ($P_{det}$) between wires 800 nm apart for Ru → Pt ($1.68 \times 10^{-15} \pm 1.50 \times 10^{-17}$) and Pt → Ru ($1.24 \times 10^{-15} \pm 4.58 \times 10^{-18}$). The data for other spacing between the injector and the detector are shown in the Supporting Information (S2).

Since the resistance of the detector influences the measured voltage, one should take into account the difference in electrical resistance between the injector and detector when comparing the two processes. To account for this difference between the nanowires ($R_{Pt}$ = 12.3 kΩ and $R_{Ru}$ = 29.3 kΩ), the current is adjusted so that the electrical power applied to both injectors is the same in both cases (1.1 mW). Here, we highlight again that the signal detected at the detector in our experiments solely arises from the magnons polarized by the injector. The voltage signal arising from thermal magnons is not responsible for the measured $R_{nl}$. We also emphasize that the magnitude of $R_{nl}$ differs depending on whether the magnons are generated solely by the SHE in Pt and detected by iOHE and iSHE at the Ru/Pt wire or generated by the OHE and SHE in a Ru/Pt wire and detected via iSHE in Pt.

To better understand the asymmetry in the magnon generation/detection process, we plotted $P_{det}$ as a function of $P_{inj}$ for Pt to Ru (red) and Ru to Pt (blue), as shown in Figure 3(b). The measurements are performed for different distances between the injection and detection wires and for different injected powers $P_{inj}$ for each measurement; the data is fitted to a sinusoidal function, the value of the amplitude is extracted for each $P_{inj}$ in each device, and $P_{det}(P_{inj})$ is fitted to the line $P_{det}^{fit}(P_{inj})$. The efficiency from $P_{inj}$ to $P_{det}$ is defined as

$$\xi = \frac{\partial P_{det}^{fit}(P_{inj})}{\partial P_{inj}} \quad (2)$$

These efficiencies show a clear difference ($\xi_{Pt \to Ru}$ = $1.24 \times 10^{-15} \pm 4.58 \times 10^{-18}$ compared to $\xi_{Ru \to Pt}$ = $1.68 \times 10^{-15} \pm 1.5 \times 10^{-17}$), and the respective linear fits can be seen in Figure 3 (b) for Pt → Ru (red) and Ru → Pt (blue).

We have demonstrated nonreciprocal generation and detection of magnons in YIG, attributed to the introduction of a Ru layer in one of our electrodes, by reversing the roles of injector and detector in our experiments. To explain this, we note that a similar nonreciprocity in charge-spin current mutual interconversion has been predicted and experimentally observed in NiFe/CuOx samples due to additional spin current generation via spin vorticity coupling.[43] This effect arises from a resistivity gradient along the thickness of CuOx, leading to a nonuniform spatial distribution in the drift velocity of conduction electrons. Such an inhomogeneous drift velocity can be seen as uniform motion upon which a rotation has been superimposed,[44,45] the latter being captured via the so-called vorticity.[44,45] Thus, the inhomogeneous drift induces a nonzero motional orbital angular momentum which is transferred to the spin through the spin-vorticity coupling.[44,45] Consequently, the process of charge-spin conversion has an additional vorticity contribution. In our Ru/Pt bilayers, a similar variation in the drift velocity emerges around the Pt/Ru interface due to different mobilities in the two materials. Thus, our observed nonreciprocity can be understood in an analogous fashion and is well represented by the bilayer model considered in the Supporting Information of ref 43. We further note that while our experiments observe a nonreciprocity in the spin−orbital-charge conversion, similar to ref 43., Onsager reciprocity is not violated by our results. The emergence of an additional drive - vorticity - in one of our electrodes causes our standard injection-detection measurement protocol to simultaneously demonstrate multiple elements of the Onsager response matrix, leading to an observed nonreciprocity despite a symmetric or reciprocal Onsager response matrix.

In conclusion, we have identified nonreciprocal processes in a nonlocal device with Ru that exhibits a strong OHE. We have found that the reciprocity observed in Pt → Pt, irrespective of the wire thickness and resistance, where SHE and iSHE are means of magnon generation and detection, does not hold once Ru is included. We compare the situation where the magnon generation (detection) involves both OHE and SHE (iOHE and iSHE), first by measuring the change in $R_{nl}$ while applying the same power in both Pt → Ru and Ru → Pt and then measuring the $P_{det}$ in the detector while applying different $P_{inj}$ in the injector and calculating the power to power efficiency in both directions. We obtain a strongly nonreciprocal efficiency $\xi_{Pt \to Ru}$ = 0.73 $\xi_{Ru \to Pt}$. This strong nonreciprocity is thus a salient feature when the interconversion among charge, spin, and orbital angular momentum occurs. We consider the emergence of vorticity around the Ru/Pt interface as a plausible mechanism underlying our observed nonreciprocity.

Note: We note that during the preparation of the manuscript, we became aware of a related work by J.A. Mendoza-Rodarte et al.,[46] which, however, relies on a comparison of the spin injection and the thermal spin generation as two separate effects.

## ASSOCIATED CONTENT

**Supporting Information**

The Supporting Information is available free of charge at https://pubs.acs.org/doi/10.1021/acs.nanolett.4c06056.

> Experiments on identical Pt wires and experiments on the power-to-power efficiency dependence on the distance between wires (PDF)

## AUTHOR INFORMATION

**Corresponding Authors**

Sachin Krishnia − *Institute of Physics, Johannes Gutenberg University Mainz, 55099 Mainz, Germany*;
Email: krishnia@uni-mainz.de

Mathias Kläui − *Institute of Physics, Johannes Gutenberg University Mainz, 55099 Mainz, Germany; Graduate School of Excellence Materials Science in Mainz, 55099 Mainz, Germany; Department of Physics, Center for Quantum Spintronics, Norwegian University of Science and Technology, 7491 Trondheim, Norway; Max Planck Graduate Center Mainz, 55122 Mainz, Germany;* orcid.org/0000-0002-4848-2569; Email: klaeui@uni-mainz.de

Authors

José Omar Ledesma-Martin − *Institute of Physics, Johannes Gutenberg University Mainz, 55099 Mainz, Germany; Max Planck Graduate Center Mainz, 55122 Mainz, Germany;* orcid.org/0000-0001-8421-7300

Edgar Galindez-Ruales − *Institute of Physics, Johannes Gutenberg University Mainz, 55099 Mainz, Germany*

Felix Fuhrmann − *Institute of Physics, Johannes Gutenberg University Mainz, 55099 Mainz, Germany*

Minh Duc Tran − *Institute of Physics, Johannes Gutenberg University Mainz, 55099 Mainz, Germany*

Rahul Gupta − *Institute of Physics, Johannes Gutenberg University Mainz, 55099 Mainz, Germany*

Marcel Gasser − *Institute of Physics, Johannes Gutenberg University Mainz, 55099 Mainz, Germany*

Dongwook Go − *Institute of Physics, Johannes Gutenberg University Mainz, 55099 Mainz, Germany; Peter Grünberg Institut and Institute for Advanced Simulation, Forschungszentrum Jülich and JARA, 52425 Jülich, Germany;* orcid.org/0000-0001-5740-3829

Akashdeep Kamra − *Department of Physics and Research Center OPTIMAS, Rheinland-Pfälzische Technische Universität Kaiserslautern-Landau, 67663 Kaiserslautern, Germany*

Gerhard Jakob − *Institute of Physics, Johannes Gutenberg University Mainz, 55099 Mainz, Germany; Max Planck Graduate Center Mainz, 55122 Mainz, Germany;* orcid.org/0000-0001-9466-0840

Yuriy Mokrousov − *Institute of Physics, Johannes Gutenberg University Mainz, 55099 Mainz, Germany; Peter Grünberg Institut and Institute for Advanced Simulation, Forschungszentrum Jülich and JARA, 52425 Jülich, Germany; Max Planck Graduate Center Mainz, 55122 Mainz, Germany*

Complete contact information is available at:
https://pubs.acs.org/10.1021/acs.nanolett.4c06056

Notes
The authors declare no competing financial interest.

■ ACKNOWLEDGMENTS

The authors thank the DFG (Spin+X (A01, A11, A12, B02) TRR 173-268565370 and Project No. 358671374), the Horizon 2020 Framework Programme of the European Commission under FETOpen Grant Agreement No. 863155 (s-Nebula), the European Research Council Grant Agreement No. 856538 (3D MAGiC), and the Research Council of Norway through its Centers of Excellence funding scheme, Project No. 262633 "QuSpin". The study also has been supported by the European Horizon Europe Framework Programme under an EC Grant Agreement No. 101129641 "OBELIX". This work was supported by the Max Planck Graduate Center with the Johannes Gutenberg University of Mainz (MPGC). This research is part of the TOPOCOM project, which is funded by the European Union's Horizon Europe Programme Horizon.1.2 under the Marie Skłodowska-Curie Actions (MSCA), Grant Agreement No. 101119608.